\documentclass[acmsmall]{acmart}
\pdfoutput=1

\startPage{1}
\setcopyright{none}
\renewcommand\footnotetextcopyrightpermission[1]{} 
\usepackage{url}
\usepackage{listings}
\usepackage[T1]{fontenc} 
\usepackage{graphicx}
\usepackage{float}
\usepackage{amsthm}
\usepackage{amsmath}
\usepackage{amssymb}
\usepackage{mathtools}
\usepackage{booktabs}
\usepackage{tabularx}
\usepackage{subcaption}
\usepackage{algorithm}
\usepackage{algorithmicx}
\usepackage[noend]{algpseudocode}

\newcommand{\code}[1]{{\small \ttfamily #1}}

\definecolor{kscolor}{rgb}{0.9,0.1,0.1}

\definecolor{mpcolor}{rgb}{0.1,0.9,0.1}

\newcommand{\name}[1]{DeepBugs}

\definecolor{eclipsePurple}{RGB}{166,38,164}
\definecolor{keyword-orange}{RGB}{152,104,1}
\definecolor{special-red-keyword}{RGB}{228,86,73}
\definecolor{accesspathBlue}{RGB}{1,132,188}
\definecolor{comment-color}{RGB}{92,99,112}
\definecolor{string-green}{RGB}{80,161,79}
\definecolor{thirdlevelaccesspathBlue}{RGB}{64,120,242}

\lstdefinelanguage{JavaScript}{
    classoffset=0,
    morekeywords={typeof, new, true, false, try, function, return, null, catch, switch, var, if, in, while, do, else, case, break},
    keywordstyle=\color{eclipsePurple!70!black},
    classoffset=1,
    morekeywords={prototype,class, export, boolean, throw, implements, import, this, hasOwnProperty, push, length,parse},
    keywordstyle=\color{special-red-keyword},
    classoffset=2,
    morekeywords={Number, Math, console, window,Array,Date,Object},
    keywordstyle=\color{keyword-orange},
    classoffset=3,
    morekeywords={log, pow,round,isNaN,indexOf,splice,apply,keys,eval},
    keywordstyle=\color{accesspathBlue!80!black},
    classoffset=4,
    morekeywords={round,remove},
    keywordstyle=\color{thirdlevelaccesspathBlue},
    identifierstyle=\color{black},
    sensitive=false,
    comment=[l]{//},
    morecomment=[s]{/*}{*/},
    commentstyle=\itshape\color{comment-color!80}\ttfamily,
    stringstyle=\color{string-green}\ttfamily,
    morestring=[b]',
    morestring=[b]"
}
\theoremstyle{plain}

\begin{document}

\title{\name{}:\\ A Learning Approach to Name-based Bug Detection}

\author{Michael Pradel}
\affiliation{
  \department{Department of Computer Science}
  \institution{TU Darmstadt, Germany}
}

\author{Koushik Sen}
\affiliation{
  \department{EECS Department}
  \institution{University of California, Berkeley, USA}
}

\begin{abstract}
Natural language elements in source code, e.g., the names of variables and functions, convey useful information.
However, most existing bug detection tools ignore this information and therefore miss some classes of bugs.
The few existing name-based bug detection approaches reason about names on a syntactic level and rely on manually designed and tuned algorithms to detect bugs.
This paper presents \name{}, a learning approach to name-based bug detection, which reasons about names based on a semantic representation and which automatically learns bug detectors instead of manually writing them.
We formulate bug detection as a binary classification problem and train a classifier that distinguishes correct from incorrect code.
To address the challenge that effectively learning a bug detector requires examples of both correct and incorrect code, we create likely incorrect code examples from an existing corpus of code through simple code transformations.
A novel insight learned from our work is that learning from artificially seeded bugs yields bug detectors that are effective at finding bugs in real-world code.
We implement our idea into a framework for learning-based and name-based bug detection. Three bug detectors built on top of the framework detect accidentally swapped function arguments, incorrect binary operators, and incorrect operands in binary operations.
Applying the approach to a corpus of 150,000 JavaScript files yields bug detectors that have a high accuracy (between 89\% and 95\%), are very efficient (less than 20 milliseconds per analyzed file), and reveal 102 programming mistakes (with 68\% true positive rate) in real-world code.
\end{abstract}

\maketitle

\thispagestyle{empty}

\section{Introduction}

Source code written by humans contains valuable natural language information, such as the identifier names of variables and functions.
This information often conveys insights into the semantics intended by the developer, and therefore is crucial for human program understanding~\cite{Lawrie2006,Butler2010}.
While the importance of identifier names for humans is widely recognized, program analyses typically ignore most or even all identifier names.
For example, popular static analysis tools, such as Google Error Prone~\cite{Aftandilian2012} , FindBugs~\cite{Hovemeyer2004}, or lgtm\footnote{\url{https://lgtm.com/}}, mostly ignore identifier names.
As a result, analyzing a program that has meaningful identifiers chosen by human developers yields the same results as analyzing a variant of the program where identifiers are consistently replaced with arbitrary and meaningless names.

Ignoring identifier names causes existing bug detection tools to miss bugs that, in hindsight, may appear obvious to a human.
Table~\ref{tab:exampleBugs} gives three examples of such bugs.
All three are from real-world code written in JavaScript, a language where identifiers are particularly important due to the lack of static types.
Example~1 shows a bug in Angular.js where the developer accidentally passes two function arguments in the wrong order.
The first expected argument is a callback function, but the second argument is called \code{fn}, an abbreviation for ``function''.
Example~2 shows a bug in the Angular-UI-Router project where the developer compares two values of incompatible types with each other.
In the absence of statically declared types, this inconsistency can be spotted based on the unusual combination of identifier names.
Finally, Example~3 shows a bug in the DSP.js library where the developer accidentally swapped the operands of a binary operation inside a loop.
A human might detect this bug knowing that \code{i} is a common name for a loop variable, which suggests that the code does not match the intended semantics.
As illustrated by these examples, identifier names convey valuable information that can help to detect otherwise missed programming mistakes.

\begin{table}[]
    \centering
    \caption{Examples of name-related bugs detected by \name{}.}
    \label{tab:exampleBugs}
    \renewcommand{\arraystretch}{1.4}
    \begin{tabular}{rlp{15em}}
        \toprule
        ID & Buggy code & Description \\
        \midrule
        1 &
\begin{lstlisting}
browserSingleton.startPoller(100,
  function(delay, fn) {
    setTimeout(delay,fn);
});
\end{lstlisting} &
        \vspace{-2em}
        The \code{setTimeout} function expects two arguments: a callback function and the number of milliseconds after which to invoke the callback. The code accidentally passes these arguments in the inverse order. \\
        2 &
\begin{lstlisting}
for (j = 0; j < param.replace; j++) {
  if (param.replace[j].from === paramVal)
    paramVal = param.replace[j].to;
}
\end{lstlisting} &
        \vspace{-2em}
        The header of the for-loop compares the index variable \code{j} to the array \code{param.replace}. Instead, the code should compare \code{j} to \code{param.replace.length}.\\
        3 &
\begin{lstlisting}
for(var i = 0; i<this.NR_OF_MULTIDELAYS; i++){
  // Invert the signal of every even multiDelay
  outputSamples = mixSampleBuffers(outputSamples,
      this.multiDelays[i].process(filteredSamples),
      2%i==0, this.NR_OF_MULTIDELAYS);
    /*^^^^^^*/
}
\end{lstlisting} &
        \vspace{-3.2em}
        The highlighted expression \code{2\%i==0} is supposed to alternate between \code{true} and \code{false} while traversing the loop. However, the code accidentally swapped the operands and should instead be \code{i\%2==0}.\\
        \bottomrule 
    \end{tabular}

\end{table}

One reason why most program analyses, including bug detection tools, ignore identifier names is that reasoning about them is hard.
Specifically, there are two challenges for a name-based bug detector.
First, a name-based analysis must reason about the meaning of identifier names.
As a form of natural language information, identifier names are inherently fuzzy and elude the precise reasoning that is otherwise common in program analysis.
Second, given an understanding of the meaning of identifier names, an analysis must decide whether a given piece of code is correct or incorrect.
To be practical as a bug detection tool, the second challenge must be addressed in a way that yields a reasonably low number of false positives while detecting actual bugs.

Previous work on name-based bug detection~\cite{Host2009,issta2011,icse2016-names,oopsla2017} addresses these challenges by lexically reasoning about identifiers and through manually designed algorithms.
To reason about the meaning of identifiers, these approaches use lexical similarities of names as a proxy for semantic similarity.
For example, the existing approaches may find \code{length} and \code{len} to be similar because they share a common substring, but miss the fact that \code{length} and \code{count} are semantically similar even though they are lexically different.
To make decisions about programs, e.g., to report a piece of code as likely incorrect, existing name-based analyses rely on manually designed algorithms that use hard-coded patterns and carefully tuned heuristics.
For example, a name-based analysis that has been recently deployed at Google~\cite{oopsla2017} comes with various heuristics to increase the number of detected bugs and to decrease the number of false positives.
Designing and fine-tuning such heuristics imposes a significant human effort that is difficult to reuse across different analyses and different classes of bugs.

This paper tackles the problem of name-based bug detection with a machine learning-based approach.
To address the problem of reasoning about the meaning of identifiers, we use a learned vector representation of identifiers.
This representation, called embeddings, preserves semantic similarities, such as the fact that \code{length} and \code{count} are similar.
Such embeddings have been successful for several natural language processing tasks and adopting them to source code is a natural choice for name-based bug detection.
To address the problem of deciding whether a piece of code is likely correct or incorrect, we formulate the problem as binary classification and train a model to distinguish correct from incorrect code.
Because the classifier is learned without human intervention, the approach does not rely on designing and tuning heuristics.

Effectively learning a classifier that distinguishes correct from incorrect code requires training data that consists of both correct and incorrect examples.
Examples of correct code are easily available due to the huge amounts of existing code, based in the common assumption that most parts of most code are correct.
In contrast, large amounts of code examples  that are incorrect for a specific reason are much harder to find.
In particular, manually obtaining a sufficiently large data set would require a human to label thousands of bugs.
To address this problem, we generate large amounts of training data via simple program transformations that insert likely bugs into existing, supposedly correct code.
An important insight of our work is that learning from such artificially generated training data yields a learned model that is effective at identifying real-world bugs.

We implement our ideas into an extensible framework, called \name{}, that supports different classes of name-related bugs.
The framework extracts positive training examples from a code corpus, applies a simple transformation to also create large amounts of negative training examples, trains a model to distinguish these two, and finally uses the trained model for identifying mistakes in previously unseen code.
We present three bug detectors based on \name{} that find accidentally swapped function arguments, incorrect binary operators, and incorrect operands in binary operations.
Creating a new bug detector consists of two simple steps.
First, provide a training data generator that extracts correct and incorrect code examples from a given corpus of code.
Second, map each code example into a vector that the machine learning model learns to classify as correct or incorrect.
For the second step, all bug detectors reuse the same embedding of identifier names, simplifying the task of creating a name-based bug detector.

Our approach differs from existing bug detectors that identify bugs as anomalies in a corpus of code~\cite{Hangal2002,Engler2001,Monperrus2010}.
These approaches infer information from existing code by learning only from correct examples and then flag any code as unusual that deviates from the norm.
To reduce false positives, those approaches typically filter the detected anomalies based on manually designed heuristics.
Instead, our approach learns from positive and negative examples, enabling the machine learning model to accurately distinguish these two classes.
\name{} differs from existing work on name-based bug detection~\cite{Host2009,issta2011,icse2016-names,oopsla2017} by reasoning about identifiers based on a semantic representation, by learning bug detectors instead of manually writing them, and by considering two new name-related bug patterns on top of the previously considered swapped function arguments~\cite{issta2011,icse2016-names,oopsla2017}.
Finally, our work is the first on name-based bug detection for dynamically typed languages, where name-related bugs may remain unnoticed due to the lack of static type checking.

We evaluate \name{} and its three instantiations by learning from a corpus of 100,000 JavaScript files and by searching mistakes in another 50,000 JavaScript files.
In total, the corpus amounts to 68 million lines of code.
We find that the learned bug detectors have an accuracy between 89\% and 95\%, i.e., they are very effective at distinguishing correct from incorrect code.
Manually inspecting a subset of the warnings reported by the bug detectors, we found 102 real-world bugs and code quality problems among 150 inspected.
Even though we do not perform any manual tuning or filtering of warnings, the bug detectors have a reasonable precision of 68\%, i.e., the majority of the reported warnings point to actual bugs.

In summary, this paper contributes the following:
\begin{itemize}
    \item A learning approach to name-based bug detection, which differs from previous name-based bug detectors (i) by reasoning about identifier names based on a semantic representation, (ii) by learning bug detectors instead of manually writing them, (iii) by considering additional bug patterns, and (iv) by targeting a dynamically typed programming language.
     
    \item We formulate bug detection as a classification problem and present a framework to learn a classifier from examples of correct and incorrect code. To obtain large amounts of training data for both classes, we create training data through simple program transformations that yield likely incorrect code.
    
    \item We implement the idea into a general framework that can be instantiated into different kinds of name-based bug detectors.
    The framework is available as open-source, enabling others to build on our work, e.g., by adding further bug detectors:\\ \url{https://github.com/michaelpradel/DeepBugs}
        
    \item We provide empirical evidence that the approach yields effective bug detectors that find various bugs in real-world JavaScript code.
    
\end{itemize}

\section{A Framework for Learning to Find Name-Related Bugs}
\label{sec:framework}

This section presents the \name{} framework for automatically creating name-based bug detectors via machine learning.
The basic idea is to train a classifier to distinguish between code that is an instance of a name-related bug pattern and code that does not suffer from this bug pattern.
By bug pattern, we informally mean a class of programming errors that are similar because they violate the same rule.
For example, accidentally swapping the arguments passed to a function, calling the wrong API method, or using the wrong binary operator are bug patterns.
Manually written bug checkers, such as FindBugs or Error Prone, are also based on bug patterns, each of which corresponds to a separately implemented analysis.

\subsection{Overview}

\begin{figure*}
    \centering
    \includegraphics[width=\linewidth]{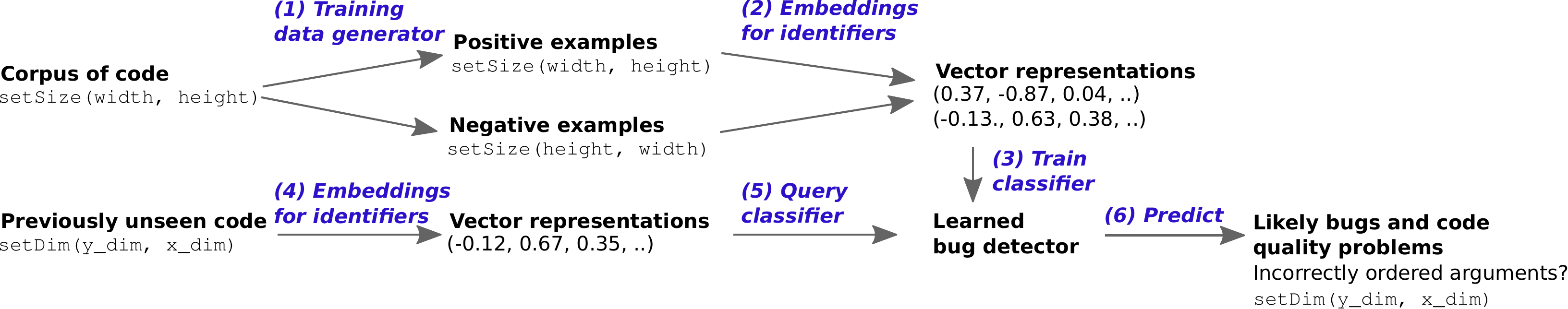}
    \caption{Overview of our approach.}
    \label{fig:overview}
\end{figure*}

Given a corpus of code, creating and using a bug detector based on \name{} consists of several steps. Figure~\ref{fig:overview} illustrates the process with a simple example.
\begin{enumerate}
\item \emph{Extract and generate training data from the corpus}.
This step statically extracts \emph{positive}, i.e., likely correct, code examples from the given corpus and generates \emph{negative}, i.e., likely incorrect, code examples.
Because we assume that most code in the corpus is correct, the extracted positive code examples are likely to not suffer from the particular bug pattern.
To also create negative training examples, \name{} applies simple code transformations that are likely to introduce a bug.
(Step~1 in Figure~\ref{fig:overview}.)

\item \emph{Represent code as vectors}.
This step translates each code example into a vector.
To preserve semantic information conveyed by identifier names, we learn an embedding that maps identifiers to a semantic vector representation via a Word2Vec neural network~\cite{Mikolov2013a}.
(Step~2 in Figure~\ref{fig:overview}.)

\item \emph{Train a model to distinguish correct and incorrect examples}.
Given two sets of code that contain positive and negative examples, respectively, this step trains a classifier to distinguish between the two kinds of examples. The classifier is a feedforward neural network.
(Step~3 in Figure~\ref{fig:overview}.)

\item \emph{Predict bugs in previously unseen code}.
This step applies the classifier obtained in the previous step to predict whether a previously unseen piece of code suffers from the bug pattern.
If the learned model classifies the code to be likely incorrect, the approach reports a warning to the developer.
(Steps~4 to~6 in Figure~\ref{fig:overview}.)
\end{enumerate}

The example in Figure~\ref{fig:overview} illustrates these steps for a bug detector aimed at finding incorrectly ordered function arguments.
In Step~6, the bug detector warns about a likely bug where the arguments \code{y\_dim} and \code{x\_dim} should be swapped.
The reason that the approach can spot such bugs is that the trained classifier generalizes beyond the training data based on the semantic representation of identifiers.
For the example, the representation encodes that \code{width} and \code{x\_dim}, as well as \code{height} and \code{y\_dim}, are pairwise semantically similar, enabling \name{} to detect the bug.

\subsection{Generating Training Data}
\label{sec:generatingData}

An important prerequisite for any learning-based approach is a sufficiently large amount of training data.
In this work, we formulate the problem of bug detection as a binary classification task and address it via supervised learning.
To effectively address this task, our approach relies on training data for both classes, i.e., examples of both correct and incorrect code.
As observed by others~\cite{Nguyen2015,Raychev2015,Bielik2016}, the huge amount of existing code provides ample of examples of likely correct code.
In contrast, it is non-trivial to obtain many examples of code that suffers from a particular bug pattern.
One possible approach is to manually or semi-automatically search code repositories and bug trackers for examples of bugs that match a given bug pattern.
However, scaling this approach to thousands or even millions of examples, as required for advanced machine learning, is extremely difficult.

Instead of relying on manual effort for creating training data, this work generates training data fully automatically from a given corpus of code.
The key idea is to apply a simple code transformation $\tau$ that transforms likely correct code extracted from the corpus into likely incorrect code.
Section~\ref{sec:instances} presents implementations of $\tau$ that apply simple AST-based code transformations.


\begin{definition}[Training data generator]
	\label{def:trainingDataGenerator}
Let $C \subseteq L$ be a set of code in a programming language $L$.
Given a piece of code $c \in C$, a training data generator $G: C \rightarrow (2^{C_{pos}}, 2^{C_{neg})}$ creates two sets of code snippets $C_{pos} \subseteq C$ and $C_{neg} \subseteq L$, which contain positive and negative training examples, respectively.
The negative examples are created by applying transformation $\tau: C \rightarrow C$ to each positive example:
$C_{neg} = \{ c_{neg} ~|~ c_{neg} = \tau(c_{pos}) \ \forall c_{pos} \in C_{pos} \}$
\end{definition}

By \emph{code snippet} we mean a single expression, a single statement, or multiple related statements.
Each code snippet contains enough information to determine whether it contains a bug.
For example, a code snippet can be a call expression with two arguments, which enables a bug detector to decide whether the arguments are passed in the correct order.

There are various ways to implement a training data generator.
For example, suppose the bugs of interest are accidentally swapped arguments of function calls.
A training data generator for this bug pattern gathers positive examples by extracting all function calls that have at least two arguments and negative examples by permuting the order of these arguments of the function calls.
Under the assumption that the given code is mostly correct, the original calls are likely correct, whereas changing the order of arguments is likely to provide an incorrect call.

Our idea of artificially creating likely incorrect code relates to mutation testing~\cite{jia2011analysis} and to work on artificially introducing security vulnerabilities~\cite{Dolan-Gavitt2016,Pewny2016}.
These existing techniques are intended for evaluating the effectiveness of test suites and of vulnerability detection tools, respectively.
Our work differs by creating likely incorrect code for a different purpose, training a machine learning model, and by considering bug patterns that are amenable to name-based bug detection.

\subsection{Embeddings for Identifiers and Literals}
\label{sec:embeddings}

As machine learning relies on vector representations of the analyzed data, to learn a bug detector, we require vector representations of code snippets.
An important challenge for a name-based bug detector, is to reason about identifier names, which are natural language information and therefore inherently difficult to understand for a computer.
Our goal is to distinguish semantically similar identifiers from dissimilar ones. 
For example, the bug detector that searches for swapped arguments may learn from examples such as \code{done(error, result)} that \code{done(res, err)} is likely to be wrong, because \code{error} $\approx$ \code{err} and \code{result} $\approx$ \code{res}, where $\approx$ refers to semantic similarity.
In contrast, \code{seq} and \code{sequoia} are semantically dissimilar because they refer to different concepts, even though they share a common prefix of characters.
As illustrated by these examples, semantic similarity does not always correspond to lexical similarity, as considered by prior work~\cite{issta2011,icse2016-names,oopsla2017}, and may even exist cross type boundaries.
To enable a machine learning-based bug detector to reason about identifiers, we require a representation of identifiers that preserves semantic similarities.

In addition to identifiers, we also consider literals in code, such as \code{true} and \code{23}, because they also convey relevant semantic information that can help to detect bugs.
For example, \code{true} and \code{1} are similar (in JavaScript, at least) because both evaluate to \code{true} when being used in a conditional.
To simplify the presentation, we say ``identifier'' to denote both identifiers and literals.
Our implementation disambiguates tokens that represent identifiers and literals from each other and from language keywords by prepending the former with ``ID:'' and the latter with ``LIT:''.

\name{} reasons about identifiers by automatically learning a vector representation, called embeddings, for each identifier based on a corpus of code:

\begin{definition}[Embeddings]
The embeddings are a map $E: I \rightarrow \mathbb{R}^e$ that assigns to each identifier in the set $I$ of identifiers a real-valued vector in an $e$-dimensional space.
\end{definition}

A na\"ive representation is a local, or one-hot, encoding, where $e=|I|$ and where each vector returned by $E$ contains only zeros except for a single element that is set to one and that represents the specific identifier.
Such a local representation fails to provide two important properties.
First, to enable efficient learning, we require an embedding that stores many identifiers in relatively short vectors.
Second, to enable \name{} to generalize across non-identical but semantically similar identifiers, we require an embedding that assigns a similar vector to semantically similar identifiers.

Instead of a local embedding, we use a distributed embedding, where the information about an identifier is distributed across all elements of the vector returned by $E$.
Our distributed embedding is inspired by word embeddings for natural languages, specifically by Word2Vec~\cite{Mikolov2013}.
The basic idea of Word2Vec is that the meaning of a word can be derived from the various contexts in which this word is used.
In natural languages, the context of an occurrence of a word in a sequence of words is the window of words preceding and succeeding the word.

We adapt this idea to source code by viewing code as a sequence of tokens and by defining the context of the occurrence of an identifier as its immediately preceding and succeeding tokens.
Given a sequence of tokens $t_1, ..., t_i, ..., t_k$, where $t_i$ is an identifier, the approach considers a window of $w$ tokens around $t_i$, containing the $\frac{w}{2}$ tokens before $t_i$ and the $\frac{w}{2}$ tokens after $t_i$.
As a default, we choose $w=20$ in our experiments.

We use the CBOW variant of Word2Vec~\cite{Mikolov2013}, which trains a neural network that predicts a token from its surrounding tokens.
To achieve this goal, the network feeds the given information through a hidden layer of size $e$, from which the token is predicted.
We use $e=200$ for our experiments.
Once trained, the network has learned a semantics-preserving representation of each identifier, and we use this representation as the embedding $E$ of the identifier.

For efficiency during training, we limit the vocabulary $V$ of tokens, including identifiers and literals, to $|V|=10,000$ by discarding the least frequent tokens.
To represent tokens beyond $V$, we use a placeholder ``unknown''.
Section~\ref{sec:evalVocab} shows that this way of bounding the vocabulary size covers the vast majority of all occurrences of identifiers and literals.

\subsection{Vector Representations of Positive and Negative Code Examples}
\label{sec:vectorRepres}

Given code snippets extracted from a corpus, our approach uses the embeddings for identifiers to represent each snippet as a vector suitable for learning:

\begin{definition}[Code representation]
	\label{def:codeRepres}
Given a code snippet $c \in C$, its code representation $v \in \mathbb{R}^n$ is an $n$-dimensional real-valued vector that contains the embeddings of all identifiers in $c$.
\end{definition}

Each bug detector built on top of the \name{} framework chooses a code representation suitable for the specific kind of code snippet (explained in detail in Section~\ref{sec:instances}).
For example, to detect bugs related to function arguments, the code representation may contain the embeddings of the function name and the arguments.

\begin{table}[tb]
    \centering
    \caption{Examples of identifier names and literals extracted for name-based bug detectors.}
    \label{tab:nameExtraction}
    \begin{tabular}{rl}
        \toprule
        Expression & Extracted name \\
        \midrule
        \code{list} & ID:list \\
        \code{23} & LIT:23 \\
        \code{this} & LIT:this \\
        \code{i++} & ID:i \\
        \code{myObject.prop} & ID:prop \\
        \code{myArray[5]} & ID:myArray \\
        \code{nextElement()} & ID:nextElement \\
        \code{db.allNames()[3]} & ID:allNames \\
        \bottomrule
    \end{tabular}
\end{table}

All bug detectors share the same technique for extracting names of expressions.
Given an AST node $n$ that represents an expression, we extract $name(n)$ as follows:
\begin{itemize}
    \item If $n$ is an identifier, return its name.
    \item If $n$ is a literal, return a string representation of its value.
    \item If $n$ is a \code{this} expression, return ``this''.
    \item If $n$ is an update expression that increments or decrements $x$, return $name(x)$.
    \item If $n$ is a member expression $base.prop$ that accesses a property, return $name(prop)$.
    \item If $n$ is a member expression $base[k]$ that accesses an array element, return $name(base)$.
    \item If $n$ is a call expression $base.callee(..)$, return $name(callee)$.
    \item For any other AST node $n$, do not extract its name.
 \end{itemize}
Table~\ref{tab:nameExtraction} gives examples of names extracted from JavaScript expressions.
We use the prefixes ``ID:'' and ``LIT:'' to distinguish identifiers and literals.
The extraction technique is similar to that used in manually created name-based bug detectors~\cite{issta2011,icse2016-names,oopsla2017}, but omits heuristics to make the extracted name suitable for a lexical comparison of names.
For example, existing techniques remove common prefixes, such as \code{get} to increase the lexical similarity between, e.g., \code{getNames} and \code{names}.
Instead, \name{} identifies semantic similarities of names through learned embeddings.

\subsection{Training and Querying a Bug Detector}
\label{sec:trainingQuerying}

Based on the vector representation of code snippets, a bug detector is a model that distinguishes between vectors that correspond to correct and incorrect code examples, respectively.

\begin{definition}[Bug detector]
A bug detector $D$ is a binary classifier $D: C \rightarrow [0,1]$ that predicts the probability that a code snippet $c \in C$ is an instance of a particular bug pattern.
\end{definition}

Training a bug detector consists of two steps.
At first, \name{} computes for each positive example $c_{pos} \in C_{pos}$ its vector representation $v_{pos} \in \mathbb{R}^n$, which yields a set $V_{pos}$ of vectors.
Likewise, the approach computes the set $V_{neg}$ from the negative examples $c_{neg} \in C_{neg}$.
Then, we train the bug detector $D$ in a supervised manner by providing two kinds of input-output pairs: $(v_{pos}, 0)$ and $(v_{neg}, 1)$.
The output of the learned model can be interpreted as the probability that the given code snippet is incorrect.
That is, the model is trained to predict that positive code examples are correct and that negative code examples are incorrect.

In principle, a bug detector can be implemented by any classification technique.
We use a feedforward neural network with an input layer of a size that depends on the code representation provided by the specific bug detector,
a single hidden layer of size 200, and an
output layer with a single element that represents the probability computed by $D$.
We apply a dropout of 0.2 to the input layer and the hidden layer.
As the loss function, we use binary cross-entropy and train the network with the RMSprop optimizer for 10 epochs with batch size 100.

Given a sufficiently large set of training data, the bug detector will generalize beyond the training examples and one can query it with previously unseen code.
To this end, \name{} extracts code snippets $C_{new}$ in the same way as extracting the positive training data.
For example, for a bug detector that identifies swapped function arguments, the approach extracts all function calls including their unmodified arguments.
Next, \name{} computes the vector representation of each example $c_{new} \in C_{new}$, which yields a set $V_{new}$.
Finally, we query the trained bug detector $D$ with every $v_{new} \in V_{new}$ and obtain for each code snippet a prediction of the probability that it is incorrect.
To report warnings about bugs the a developer, \name{} ranks all warnings by the predicted probability in descending order.
In addition, one can control the overall number of warnings by omitting all warnings with a probability below a configurable threshold.

It is important to note that bug detectors built with \name{} do not require any heuristics or manually designed filters of warnings, as commonly used in existing name-based bug detectors~\cite{issta2011,icse2016-names,oopsla2017}.
For example, the start-of-the-art bug detector to detect accidentally swapped function arguments relies on a hard-coded list of function names for which swapping the arguments is expected, such as \code{flip}, \code{transpose}, or \code{reverse}~\cite{oopsla2017}.
Instead of hard-coding such heuristics, which is time-consuming and likely incomplete, learned name-based bug detectors infer these kinds of exceptions from the training data.

\section{Name-Based Bug Detectors}
\label{sec:instances}

This section presents three examples of name-based bug detectors built on top of the \name{} framework.
The bug detectors address a diverse set of programming mistakes: accidentally 
swapped function arguments, incorrect binary operators, and incorrect operands in binary expressions.
While the first bug pattern has been the target of previous work for statically typed languages~\cite{issta2011,icse2016-names,oopsla2017}, we are not aware of a name-based bug detector for the other two bug patterns.
Implementing new bug detectors is straightforward, and we envision future work to create more instances of our framework, e.g., based on bug patterns mined from version histories~\cite{DBLP:conf/sigsoft/HanamBM16,Brown2017a}.

Each bug detector consists of two simple ingredients.
\begin{itemize}
    \item \emph{Training data generator}.
    A training data generator that traverses the code corpus and extracts positive and negative code examples for the particular bug pattern based on a code transformation (Definition~\ref{def:trainingDataGenerator}).
    We find a simple AST-based traversal and transformation to be sufficient for all studied bug patterns.
    \item \emph{Code representation}.
    A mapping of each code example into a vector that the machine learning model learns to classify as either benign or buggy (Definition~\ref{def:codeRepres}).
    All bug detectors presented here build on the same embeddings of identifier names, allowing us to amortize the one-time effort of learning an embedding across different bug detectors.
\end{itemize}
Given these two ingredients and a corpus of training code, our framework learns a bug detector that identifies programming mistakes in previously unseen code.

The remainder of this section presents three bug detectors build on top of \name{}.

\subsection{Swapped Function Arguments}
\label{sec:swappedFunctionArguments}

The first bug detector addresses accidentally swapped arguments.
This kind of mistake can occur both in statically typed and dynamically typed languages.
For statically typed languages, this kind of bug occurs for methods that accept multiple equally typed arguments.
For dynamically typed languages, the problem is potentially more widespread because all calls that pass two or more arguments are susceptible to the mistake due to the lack of static type checking.
Example~1 in Table~\ref{tab:exampleBugs} shows a real-world example of this bug pattern.

\paragraph{Training Data Generator}

To create training examples from given code, the approach traverses the AST of each file in the code corpus and visits each call site that has two or more arguments.
For each such call site, the approach extracts the following information:
\begin{itemize}
    \item The name $n_{callee}$ of the called function.
    \item The names $n_{arg1}$ and $n_{arg2}$ of the first and second argument.
    \item The name $n_{base}$ of the base object if the call is a method call, or an empty string otherwise.
    \item The types $t_{arg1}$ and $t_{arg2}$ of the first and second argument for arguments that are literals, or empty strings otherwise.
    \item The names $n_{param1}$ and $n_{param2}$ of the formal parameters of the called function, or empty strings if unavailable.
\end{itemize}
All names are extracted using the $name$ function defined in Section~\ref{sec:vectorRepres}.
We resolve function calls heuristically, as sound static call resolution is non-trivial in JavaScript.
If either $n_{callee}$, $n_{arg1}$, or $n_{arg2}$ are unavailable, e.g., because the $name$ function cannot extract the name of a complex expression, then the approach ignores this call site.

From the extracted information, the training data generator creates for each call site a positive example
$$x_{pos} =\allowbreak (n_{base},\allowbreak n_{callee},\allowbreak n_{arg1},\allowbreak n_{arg2},\allowbreak t_{arg1},\allowbreak t_{arg2},\allowbreak n_{param1},\allowbreak n_{param2})$$
and a negative example
$$x_{neg} =\allowbreak (n_{base},\allowbreak n_{callee},\allowbreak n_{arg2},\allowbreak n_{arg1},\allowbreak t_{arg2},\allowbreak t_{arg1},\allowbreak n_{param1},\allowbreak n_{param2}).$$
That is, to create the negative example, we simply swap the arguments w.r.t.\ the order in the original code.

\paragraph{Code representation}

To enable \name{} to learn from the positive and negative examples, we transform $x_{pos}$ and $x_{neg}$ from tuples of strings into vectors.
To this end, the approach represents each string in the tuple $x_{pos}$ or $x_{neg}$ as a vector.
Each name $n$ is represented as $E(n)$, where $E$ is the learned embedding from Section~\ref{sec:embeddings}.
To represent type names as vectors, we define a function $T$ that maps each built-in type in JavaScript to a randomly chosen binary vector of length $5$.
For example, the type ``string'' may be represented by a vector $T(\text{string}) = [0,1,1,0,0]$, whereas the type ``number'' may be represented by a vector $T(\text{number}) = [1,0,1,1,0]$.
Finally, based on the vector representation of each element in the tuple $x_{pos}$ or $x_{neg}$, we compute the code representation for $x_{pos}$ or $x_{neg}$ as the concatenation the individual vectors.

\subsection{Wrong Binary Operator}
\label{sec:wrongBinaryOperator}

The next two bug detectors address mistakes related to binary operations.
At first, we consider code that accidentally uses the wrong binary operator, e.g., \code{i <= length} instead of \code{i < length}.
Such mistakes are hard to find, especially in a dynamically typed language, but identifier names can provide valuable hints when search these mistakes.

\paragraph{Training Data Generator}

The training data generator traverses the AST of each file in the code corpus and extracts the following information from each binary operation:
\begin{itemize}
    \item The names $n_{left}$ and $n_{right}$ of the left and right operand.
    \item The operator $op$ of the binary operation.
    \item The types $t_{left}$ and $t_{right}$ of the left and right operand if they are literals, or empty strings otherwise.
    \item The kind of AST node $k_{parent}$ and $k_{grandP}$ of the parent and grand-parent nodes of the AST node that represents the binary operation.
\end{itemize}
We extract the (grand-)parent nodes to provide some context about the binary operation to \name{}, e.g., whether the operation is part of a conditional or an assignment.
If either $n_{left}$ or $n_{right}$ are unavailable, then we ignore the binary operation.

From the extracted information, the approach creates a positive and a negative example:
$$x_{pos} =\allowbreak (n_{left},\allowbreak n_{right}\allowbreak, op,\allowbreak t_{left},\allowbreak t_{right},\allowbreak k_{parent},\allowbreak k_{grandP})$$
$$x_{neg} =\allowbreak (n_{left},\allowbreak n_{right},\allowbreak op',\allowbreak t_{left},\allowbreak t_{right},\allowbreak k_{parent},\allowbreak k_{grandP})$$
The operator $op' \neq op$ is a randomly selected binary operator different from the original operator.
For example, given a binary expression \code{i <= length}, the approach may create a negative example \code{i < length} or \code{i \% length}, which is likely to create incorrect code.

\paragraph{Code representation}

Similar to the above bug detector, we create a vector representation of each positive and negative example by mapping each string in the tuple to a vector and by concatenating the resulting vectors.
To map a kind of AST node $k$ to a vector, we use a map $K$ that assigns to each kind of AST node in JavaScript a randomly chosen binary vector of length $8$.







\subsection{Wrong Operand in Binary Operation}

The final bug detector addresses code that accidentally uses an incorrect operand in a binary operation.
The intuition is that identifier names help to decide whether an operand fits another given operand and a given binary operator.
For example, the bug detector may identify the \code{x} operand in \mbox{\code{height - x}} as possibly buggy because the operation was intended to be \mbox{\code{height - y}}.

\paragraph{Training Data Generator}

The training data generator extracts the same information as in Section~\ref{sec:wrongBinaryOperator}, and then replaces one of the operands with a randomly selected alternative.
That is, the positive example is
$$x_{pos} =\allowbreak (n_{left},\allowbreak n_{right},\allowbreak op,\allowbreak t_{left},\allowbreak t_{right},\allowbreak k_{parent},\allowbreak k_{grandP})$$
whereas the negative example is either 
$$x_{neg} =\allowbreak (n_{left}',\allowbreak n_{right},\allowbreak op,\allowbreak t_{left}',\allowbreak t_{right},\allowbreak k_{parent},\allowbreak k_{grandP})$$
or 
$$x_{neg} =\allowbreak (n_{left},\allowbreak n_{right}',\allowbreak op,\allowbreak t_{left},\allowbreak t_{right}',\allowbreak k_{parent},\allowbreak k_{grandP}).$$
The name and type $n_{left}'$ and $t_{left}'$ (or $n_{right}'$ and $t_{right}'$) are different from those in the positive example.
To create negative examples that a programmer might also create by accident, we use alternative operands that occur in the same file as the binary operation.
For example, given \code{bits << 2}, the approach may transform it into a negative example \code{bits << next}, which is likely to yield incorrect code.

\paragraph{Code representation}

The vector representation of the positive and negative examples is the same as in Section~\ref{sec:wrongBinaryOperator}.

\section{Implementation}

The code extraction and generation of training examples is implemented as simple AST traversals based on the Acorn JavaScript parser.\footnote{\url{https://github.com/ternjs/acorn}}
The training data generator writes all extracted data into text files.
These files are then read by the implementation of the bug detector, which builds upon the TensorFlow and Keras frameworks for deep learning.\footnote{\url{https://www.tensorflow.org/} and \url{https://keras.io/}}
The large majority of our implementation is in the generic framework, whereas the individual bug detectors are implemented in about 100 lines of code each.

\section{Evaluation}
\label{sec:evaluation}

We evaluate \name{} by applying it to a large corpus of JavaScript code.
Our main research questions are:
\begin{itemize}
	\item How effective is the approach at distinguishing correct from incorrect code?
	\item Does the approach find bugs in production JavaScript code?
	\item How long does it take to train a model and, once a model has been trained, to predict bugs?
	\item How useful are the learned embeddings of identifiers compared to a simpler vector representation?
\end{itemize}

\subsection{Experimental Setup}

As a corpus of code, we use 150,000 JavaScript files provided by the authors of earlier work~\cite{Raychev2016}.
The corpus contains files collected from various open-source projects and has been cleaned by removing duplicate files.
In total, the corpus contains 68.6 million lines of code.
We use 100,000 files for training and the remaining 50,000 files for validation.
All experiments are performed on a single machine with 48 Intel Xeon E5-2650 CPU cores, 64GB of
memory, and a single NVIDIA Tesla P100 GPU.

\subsection{Extraction and Generation of Training Data}

\begin{table}[]
    \centering
    \setlength{\tabcolsep}{7pt}
    \caption{Statistics on extraction and generation of training data.}
    \label{tab:dataTraining}
    \begin{tabular}{lrr}
        \toprule
        Bug detector & \multicolumn{2}{c}{Examples} \\
        \cmidrule{2-3}
        & Training & Validation\\
        \midrule
        Swapped arguments & 1,450,932 & 739,188 \\
        Wrong binary operator & 4,901,356 & 2,322,190  \\
        Wrong binary operand & 4,899,206 & 2,321,586 \\
        \bottomrule
    \end{tabular}
\end{table}

Table~\ref{tab:dataTraining} summarizes the training and validation data that \name{} extracts and generates for the three bug detectors.
Each bug detector learns from several millions of examples, which is sufficient for effective learning.
Half of the examples are positive and negative code examples, respectively.
Manually creating this amount of training data, including negative examples, would be impractical, showing the benefit of our automated data generation approach.

\subsection{Warnings in Real-World Code}
\label{sec:evalRealBugs}

To evaluate the effectiveness of bug detectors built with \name{}, we conduct two sets of experiments.
First, reported in the following, we apply the bug detectors to unmodified real-world code and manually inspect the reported warnings to assess the precision of the learned bug detectors.
Second, reported in Section~\ref{sec:evalSeededBugs}, we conduct a large-scale evaluation with hundreds of thousands of artificially created bugs, which allows us to study the accuracy, recall, and precision of each learned bug detector.

To study the effectiveness of \name{} on real-world code, we train each bug detector with the 100,000 training files, then apply the trained bug detector to the 50,000 validation files, and finally inspect code locations that the bug detectors report as potentially incorrect.
For the manual inspection, we sort all reported warnings by the probability that the code as incorrect, as reported by the classifier, and then inspect the top 50 warnings per bug detector.
Based on the inspection, we classify each warning in one of three categories:
\begin{itemize}
	\item \emph{Bug}. A warning points to a \emph{bug} if the code is incorrect in the sense that it does not result in the expected runtime behavior.
	
	\item \emph{Code quality problem}. A warning points to a \emph{code quality problem} if the code yields the expected runtime behavior but should nevertheless be changed to be less error-prone or be more efficient.
	This category includes code that violates widely accepted conventions and programming rules traditionally checked by static linting tools.
	Our learned bug detectors find code quality problems because such problems often correspond to unusual code that diverges from a common coding idiom.
	
	\item \emph{False positive}. A warning is a \emph{false positive} in all other cases.
	If we are unsure about the intended behavior of a particular code location, we conservatively count it as a false positive.
	We also encountered various code examples with misleading identifier names, which we classify as false positives because the decision whether an identifier is misleading is rather subjective.
	
\end{itemize}

\begin{table}
	\setlength{\tabcolsep}{1pt}
	\centering
	\caption{Results of inspecting and classifying warnings in real-world code.}
	\label{tab:inspected}
	\begin{tabular}{lrrrr}
		\toprule
		Bug detector & Reported & Bugs & Code quality  & False \\
		&&& problem & positives \\
		\midrule
		Swapped arguments & 50 & 23 & 0 & 27 \\
		Wrong binary operator & 50 & 37 & 7 & 6 \\ 
		Wrong binary operand & 50 & 35 & 0 & 15 \\
		\midrule
		Total & 150 & 95 & 7& 48 \\
		\bottomrule 
	\end{tabular}

\end{table}

Table~\ref{tab:inspected} summarizes the results of inspecting and classifying warnings.
Out of the 150 inspected warnings, 95 warnings point to bugs and 7 warnings point to a code quality problem, i.e., 68\% of all warnings point to an actual problem.
Existing manually created bug detectors typically provide similar true positives rates, but heavily rely on heuristics to filter likely false positives.
We conclude that our learned name-based bug detectors are effective at finding real-world bugs while providing high precision.

\subsubsection{Examples of Bugs}

Many of the detected problems are difficult to detect with a traditional, name-unaware analysis, because the programming mistake is obvious only when understanding the intended semantics of the involved identifiers and literals.
The three motivating examples in Table~\ref{tab:exampleBugs} are bugs detected by \name{}.
We discuss three more representative examples in the following.

\paragraph{Buggy call of \,\code{Promise.done}}
The following code is from Apigee's JavaScript SDK.
The \code{Promise.done} function expects an error followed by a result, but the second call of \code{done} passes the arguments the other way around.
\begin{lstlisting}
var p = new Promise();
if (promises === null || promises.length === 0) {
  p.done(error, result)
} else {
  promises[0](error, result).then(function(res, err) {
    p.done(res, err);/*#\label{line:doneBug}#*/
  });
}
\end{lstlisting}

\paragraph{Meaningless binary operation}
The following code from \emph{RequireJS} is a bug that was found and fixed independently of us.\footnote{\url{https://github.com/requirejs/requirejs/issues/700}}
The binary operation \code{'-' === 0} is obviously meaningless, as it always returns \code{false}.
Instead, the developer intended to write \code{indexOf('-') === 0}.
\name{} detects this problem because it handles literals the same way as identifiers.	
\begin{lstlisting}
if (args[0] && args[0].indexOf('-' === 0)) { 
  args = args.slice(1); 
} 
\end{lstlisting}

\paragraph{Buggy call of \code{assertEquals}}
The following code is from the test suite of the \emph{Google Closure} library.
The arguments of \code{assertEquals} are supposed to be the expected and the actual value of a test outcome.
Swapping the argument leads to an incorrect error message when the test fails, which makes debugging unnecessarily hard.
Google developers consider this kind of mistake a bug~\cite{oopsla2017}.
\begin{lstlisting}
assertEquals(tree.remove('merry'), null);
\end{lstlisting}


\subsubsection{Examples of Code Quality Problems}

\paragraph{Error-prone termination condition}
The following code from \emph{Phaser.js} is correct but using \code{!==}, instead of \code{<}, as the termination condition of a for-loop is generally discouraged.
The reason is that the loop risks to run out-of-bounds when the counter is incremented by more than one or assigned an out-of-bounds value, e.g., by an accidental assignment in the loop body.
\name{} finds this problem because most code follows the convention of using \code{<} in the termination condition.
\begin{lstlisting}
for (var i = 0, len = b.length; i !== len; ++i) {
  ...
}
\end{lstlisting}

\paragraph{Error-prone and inefficient logical OR}
The following code from the \emph{promiscuous} library uses the bitwise OR operator \code{|} to compare two boolean values.
Unfortunately, using bitwise instead of logical operators for booleans is known to be inefficient\footnote{\url{https://jsperf.com/bitwise-logical-and}} and may cause unexpected behavior for non-boolean operands that are supposed to be coerced into booleans.
\name{} reports this problem because the \code{is} identifiers suggest two boolean return values, which are usually compared through a logical operation, such as \code{||}.
\begin{lstlisting}
transform = value && (is(obj, value) | is(func, value)) && value.then;
\end{lstlisting}

\subsubsection{Examples of False Positives}

We discuss some representative examples of false positives.
Some of them are due to wrapper functions, e.g., \code{Math.max}, for which our approach extracts a generic name, ``max'', that does not convey the specific meaning of the value returned by the call expression.
The main reason for false positives, though, are poorly named variable names that lead to surprisingly looking but correct code.
For example, \name{} falsely reports a warning claiming that the \code{each} operand in the following statement is wrong:
\begin{lstlisting}
var cw = cs[i].width + each;
\end{lstlisting}
A close inspection of the code reveals that the operand is correct, despite the fact that the name \code{each} may not be an ideal choice for a number added to a width.
Another example is the following comparison of \code{value.length} and \code{is}:
\begin{lstlisting}
if (v.isNumber(is) && value.length !== is) {
  ..
}
\end{lstlisting}
Our ``wrong binary operand'' bug detector reports a warning about the \code{is} operand, because this name is typically used for booleans and comparing a boolean to a value stored in \code{length} is very unusual.
Again, close inspection of the code shows it to be correct, though.
These false positives illustrate an inherent limitation of name-based bug detection:
Because the approach relies in reasonable names that are in line with common practice, it may fail when developers choose to diverge from the norm.
Overall, we find these cases to be rare enough to yield a bug detector with a precision of 68\%.

\subsection{Accuracy and Recall of Bug Detectors}
\label{sec:evalSeededBugs}

In the following, we evaluate the accuracy and recall of each bug detector at a large scale based on automatically seeded bugs.
This part of our evaluation complements the manual inspection of potential real-world bugs (Section~\ref{sec:evalRealBugs}) with an automated assessment based on a large number of artificially seeded bugs.
Informally, accuracy here means how many of all the classification decisions that the bug detector makes are correct.
Recall means how many of all bugs in a corpus of code that the bug detector finds.
To evaluate these metrics, we train each bug detector on the 100,000 training files and then apply it to the 50,000 validation files.
For the validation files, we use the training data generator to extract correct code examples $C_{pos}$ and to artificially create likely incorrect code examples $C_{neg}$.
We then query the bug detector $D$ with each example $c$, which yields a probability $D(c)$ that the example is buggy.
Finally, we compute the accuracy is as follows:
$$accuracy = \frac{|\{ c ~|~ c \in C_{pos} \wedge D(c) < 0.5 \}| + |\{ c ~|~ c \in C_{neg} \wedge D(c) \geq 0.5 \}|}{|C_{pos}| + |C_{neg}|}$$

\begin{table}[]
    \centering
    \setlength{\tabcolsep}{6pt}
    \caption{Accuracy of the bug detectors with random and learned embeddings for identifiers.}
    \label{tab:accuracy}
    \begin{tabular}{lrr}
        \toprule
        & \multicolumn{2}{c}{Embedding} \\
        \cmidrule{2-3}
        & Random & Learned \\
        \midrule
        Swapped arguments & 93.88\% & 94.70\% \\
        Wrong binary operator & 89.15\% & 92.21\% \\
        Wrong binary operand & 84.79\% & 89.06\% \\
        \bottomrule
    \end{tabular}

\end{table}

The last column of Table~\ref{tab:accuracy} shows the accuracy of the bug detectors.
The accuracy ranges between 89.06\% and 94.70\%, i.e., all bug detectors are highly effective at distinguishing correct from incorrect code examples.

The recall of a bug detector is influenced by how many warnings the detector reports.
More warnings are likely to reveal more bugs, thus increasing recall, but are also more likely to report false positives. Moreover, in practice, developers are only willing to inspect some number of warnings.
To measure recall, we assume that a developer inspects all warnings where the probability $D(c)$ is above some threshold.
We model this process by turning the bug detector $D$ into a boolean function:
$$D_t(c) = \left \{ \begin{array}{rcl}
1 & \text{if} & D(c) > t\\
0 & \text{if} & D(c) \leq t\\
\end{array} \right.$$
where $t$ is a configurable threshold that controls how many warnings to report.
Based on $D_t$, we compute recall as follows:
$$recall = \frac{|\{ c ~|~ c \in C_{neg} \wedge D_t(c)=1 \}|}{|C_{neg}|}$$

Furthermore, we measure the number of false positives as follows:
$$\#\mathit{fps} = |\{ c ~|~ c \in C_{pos} \wedge D_t(c)=1 \}|$$

Note that both recall and the number of false positives are estimates based exclusively on artificially seeded bugs.
The recall measure assumes that all seeded bugs are actual bugs that should be detected.
The false positive measure assumes that all of the original code is correct and that each warning in $C_{pos}$ should therefore be counted as a false positive.
We consider these measures here because they allow us to evaluate \name{} with hundreds of thousands of artificial bugs, which complements the evaluation with real-world bugs in Section~\ref{sec:evalRealBugs}.

\begin{figure*}[tb]
    \centering
    \includegraphics[width=0.4\linewidth]{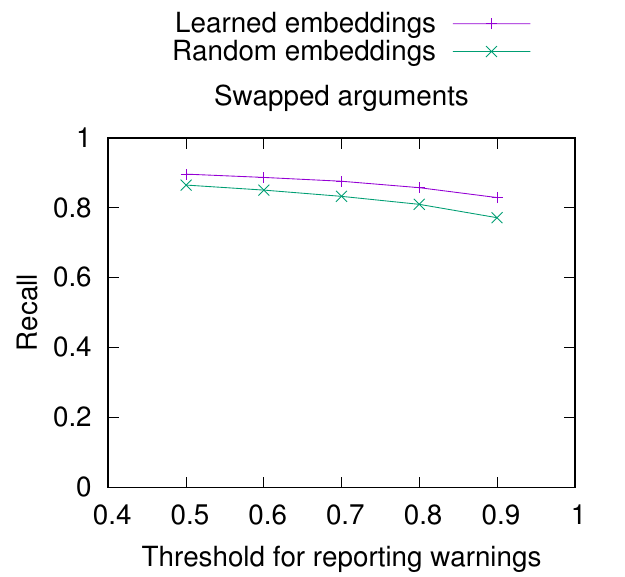}
    \includegraphics[width=0.4\linewidth]{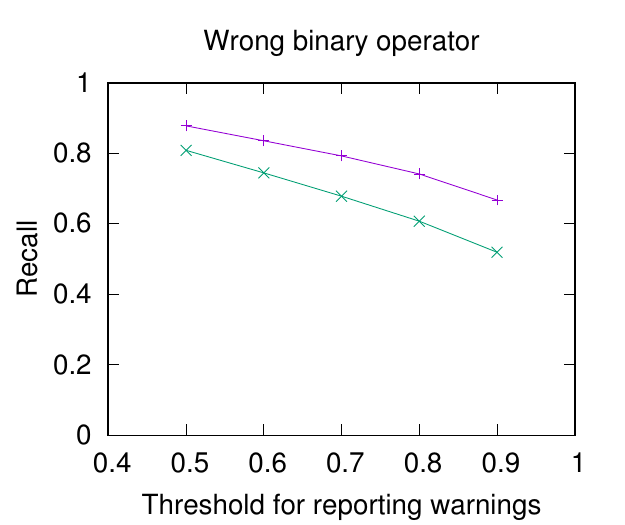}
    
    \includegraphics[width=0.4\linewidth]{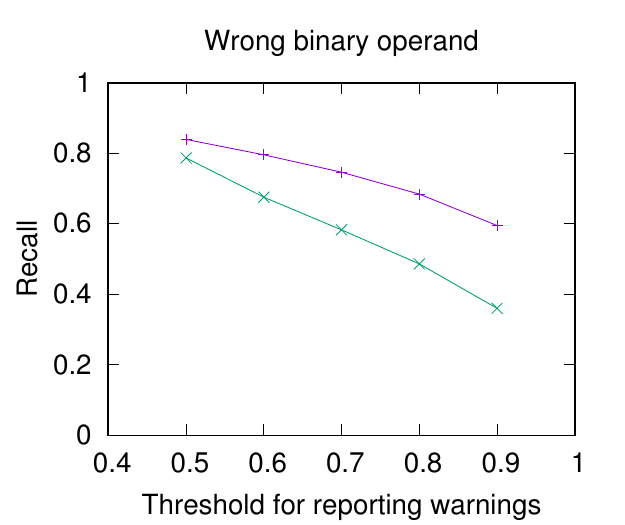}
    
    \caption{Recall of the bug detectors with different thresholds $t$ for reporting warnings. Each plot contains data points obtained with $t \in \{ 0.5, ..., 0.9 \}$. The data labeled ``Learned embedding'' corresponds to the \name{} approach.}
    \label{fig:accuracyRecall}
\end{figure*}

Figure~\ref{fig:accuracyRecall} shows the recall of the three bug detectors as a function of the threshold for reporting warnings.
Each plot shows the results for five different thresholds: $t \in \{0.5, ..., 0.9\}$.
As expected, the recall decreases when the threshold increases, because fewer warnings are reported and therefore some bugs are missed.
The results also show that some bug detectors are more likely than others to detect a bug, if a bug is present.

Considering more warnings naturally also leads to more false positives.
For a threshold of $t = 0.5$, the three bug detectors report a total of 116,941 warnings, which corresponds to roughly one warning per 196 lines of code.
In contrast, for a threshold of $t=0.9$, the number reduces to 11,292, i.e., roughly one warning per 2,025 lines of code.
In practice, we expect developers to inspect only the top-ranked warnings, as we do in Section~\ref{sec:evalRealBugs}.

%

\subsection{Efficiency}

Table~\ref{tab:times} shows how long it takes to train a bug detector and to use it to predict bugs in previously unseen code.
The training time consists of the time to gather code examples and of time to train the classifier.
The prediction time also consist of the time to extract code examples and of the time to query the classifier with each example.
Running both training and prediction on all 150,000 files takes between 36 minutes and 73 minutes per bug detector.
The average prediction time per JavaScript file is below 20 milliseconds.
We conclude that the training time is reasonable for a machine learning approach and that, once a bug detector is trained, using it on new code takes very little time.

\begin{table}[]
	\centering
	\caption{Time (min:sec) required for training and using a bug detector across the entire code corpus.}
	\label{tab:times}
	\begin{tabular}{lrr|rr}
		\toprule
		Bug detector & \multicolumn{2}{c}{Training} & \multicolumn{2}{c}{Prediction} \\
		\cmidrule{2-5}
		& Extract & Learn & Extract & Predict \\
		\midrule
		Swapped arguments & 7:46 & 20:48 & 2:56 & 5:10 \\     
		Wrong bin.\ operator & 2:44 & 49:47 & 1:28 & 8:39 \\ 
		Wrong bin.\ operand & 2:44 & 56:35 & 1:28 & 12:14 \\  
		\bottomrule
	\end{tabular}
	
\end{table}

\subsection{Usefulness of Embeddings}
\label{sec:evalEmbeddings}

Our work distinguishes itself from previous name-based bug detectors by reasoning about identifiers based on a semantic representation, i.e., the learned embeddings (Section~\ref{sec:embeddings}).
We evaluate the usefulness of learned embeddings both quantitatively and qualitatively.

For the quantitative evaluation, we compare \name{} with learned embeddings to a simpler variant of the approach that uses a baseline vector representation.
The baseline assigns to each identifier considered by \name{} a unique, randomly chosen binary vector of length $e$, i.e., the same length as our learned embeddings.
We compare the learned embeddings with the baseline w.r.t.\ accuracy and recall.
Table~\ref{tab:accuracy} shows in the ``Random'' column what accuracy the bug detectors achieve with the baseline.
Compared to the accuracy with the random embeddings, the learned embeddings yield a more accurate classifier.

Figure~\ref{fig:accuracyRecall} compares the recall of the bug detectors with the two embeddings.
We find that, for all three bug detectors, the learned embeddings increase the recall.
The reason is that the learned embeddings enable a bug detector to reason about semantic similarities between syntactically different code examples, which enables it to learn and predict bugs across similar examples.
We conclude from these results that the learned embeddings improve the effectiveness of \name{}.
At the same time, the bug detectors achieve relatively high accuracy and recall even with randomly created embeddings, showing that the overall approach has value even when no learned embeddings are available.

\begin{figure}
	\centering

\begin{minipage}{.3\linewidth}
\begin{tabular}{rl}
	\toprule
	\multicolumn{2}{c}{\emph{\textbf{ID:name}}} \\
	\midrule
	Simil. & Identifier \\
	\midrule
	0.4 & ID:names \\
	0.4 & ID:getName \\
	0.39 & ID:\_name \\
	0.39 & LIT:Identifier \\
	0.37 & ID:fullName \\
	0.36 & ID:property \\
	0.35 & ID:type \\
	0.34 & LIT::  \\
	0.34 & ID:Identifier \\
	0.34 & ID:namespace \\
	\bottomrule
\end{tabular}
\end{minipage}
\begin{minipage}{.3\linewidth}
\begin{tabular}{rl}
	\toprule
	\multicolumn{2}{c}{\emph{\textbf{ID:wrapper}}} \\
	\midrule
	Simil. & Identifier \\
	\midrule
	0.36 & ID:wrap \\
	0.36 & ID:\_wrapped \\
	0.36 & ID:wrapInner \\
	0.34 & ID:element \\
	0.32 & ID:container \\
	0.32 & ID:wrapAll \\
	0.31 & ID:attribs \\
	0.31 & ID:\$wrapper \\
	0.3 & LIT:rect \\
	0.3 & ID:renderer \\
	\bottomrule
	\end{tabular}
\end{minipage}
\begin{minipage}{.3\linewidth}
\begin{tabular}{rl}
\toprule
\multicolumn{2}{c}{\emph{\textbf{ID:msg}}} \\
\midrule
Simil. & Identifier \\
\midrule
0.57 & ID:message \\
0.46 & ID:error \\
0.4 & LIT:error \\
0.39 & ID:receive \\
0.39 & LIT:msg \\
0.39 & LIT:Error \\
0.36 & ID:alert \\
0.36 & LIT::  \\
0.35 & LIT:message \\
0.34 & LIT:log \\
\bottomrule
\end{tabular}
\end{minipage}
	
	\caption{Most similar identifiers according to the learned embeddings.}
	\label{fig:similarIds}
\end{figure}

For a qualitative assessment of the embeddings, we show for a set of identifiers which other identifiers are the most similar according to the learned embeddings.
Figure~\ref{fig:similarIds} shows the ten most similar identifiers for \code{name}, \code{wrapper}, and \code{msg}.
Recall that the ``ID:'' and ``LIT:'' prefixes are for distinguishing identifiers from literals.
The similarity scores range between 0 and 1, where 1 means the maximum similarity.

The examples illustrate three observations.
First, the embeddings encode similarities between abbreviated and non-abbreviated identifiers, e.g., \code{msg} and \code{message}.
This property is useful because abbreviations are common in source code and often introduce words not found in natural language documents.
Second, the embeddings encode similarities between lexically similar identifiers, e.g., \code{name} and \code{getName}.
Previous work on name-based bug detection relies on traditional string similarity metrics, e.g,. Levenshtein distance, and on heuristics, such as removing the \code{get} prefix, to capture these similarities.
Instead, we here learn the similarities from existing code.
Finally and perhaps most importantly, the embeddings encode semantic similarities between lexically dissimilar identifiers, such as \code{name} and \code{Identifier}, or \code{wrapper} and \code{container}.
Overall, we conclude that learned embeddings encode at least some semantic similarities of identifiers, which helps understand why a learning approach to name-based bug detection is effective.

\subsection{Vocabulary Size}
\label{sec:evalVocab}

\begin{figure}
	\centering
	\includegraphics[scale=0.8]{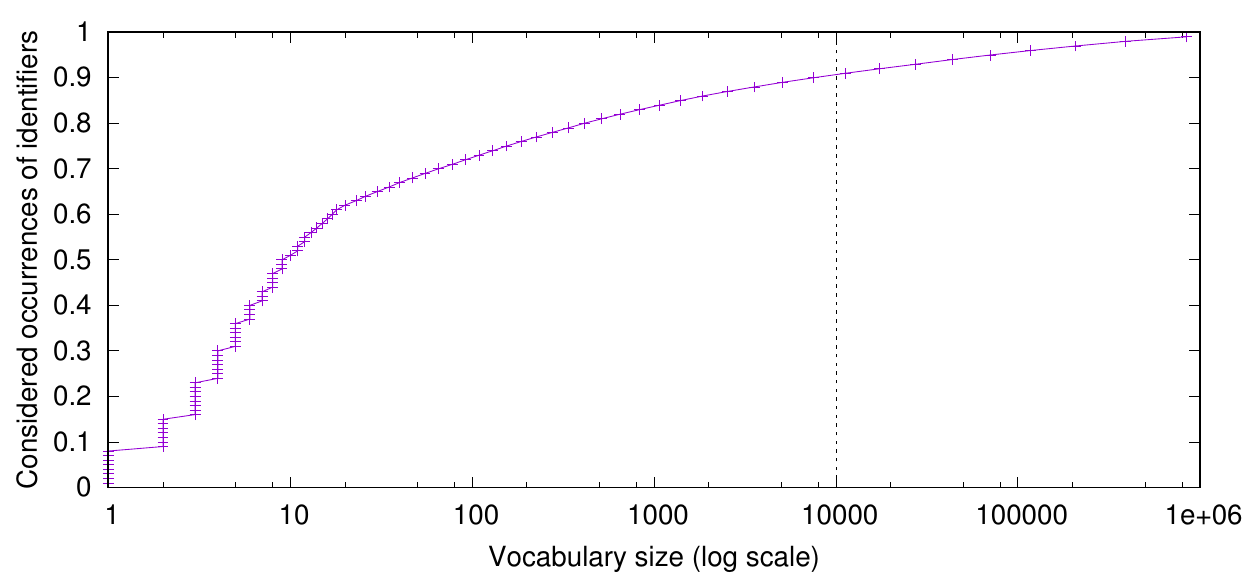}
	\caption{Number of considered identifier occurrences depending on vocabulary size.}
	\label{fig:vocab}
\end{figure}

An important parameter of our approach is the size $|V|$ of the vocabulary of tokens considered for learning embeddings (Section~\ref{sec:embeddings}).
Considering a larger vocabulary enables \name{} to reason about more code locations but also increases the resource requirements for learning the embeddings and risks to reduce accuracy due to very uncommon identifiers for which not enough training data is available.
The total number of unique tokens in the training data set is about 2.4 million, almost all of which are either identifiers or literals.
For our experiments, we set $|V|=10,000$, i.e., we consider the 10,000 most frequent tokens and replace all others with an ``unknown'' placeholder.
The reason why such a relatively small vocabulary size is effective is illustrated in Figure~\ref{fig:vocab}.
For a given vocabulary size, the figure shows the number of occurrences of identifiers that are included in the vocabulary.
Note that the horizontal axis uses a logarithmic scale.
The figure shows that a small number of tokens covers a large percentage of the occurrences of tokens.
The reason is that some tokens are very frequent, while others occur extremely rarely.
The dashed line is our default vocabulary size, which covers over 90\% of all occurrences of tokens.

\section{Related Work}


\subsection{Machine learning and Language Models for Code}

The recent successes in machine learning have lead to a strong interest in applying machine learning techniques to programs.
Existing approaches address a variety of development tasks, including
code completion based on probabilistic models of code~\cite{Raychev2014,Raychev2016a, Bielik2016},
predicting fixes of compilation errors via neural networks~\cite{Bhatia2016,Gupta2017},
and the generation of inputs for fuzz testing via probabilistic, generative language models~\cite{Patra2016} or neural networks~\cite{Godefroid2017,Liu2017,Amodio2017}.
Deep neural networks also help recommend API usages~\cite{Gu2016},
detect code clones~\cite{White2016},
classify programs into pre-defined categories~\cite{Mou2016},
and adapt copied-and-pasted code to its surrounding code~\cite{Allamanis2017}.
A recent survey discusses more approaches~\cite{Allamanis2017a}.
All these approaches exploit the availability of a large number of examples 
to learn from, e.g., in the form of publicly available code repositories, 
and that source code has regularities even across projects written by 
different developers~\cite{Hindle2012}.
We also exploit this observation but for a different tasks, bug finding, than the above work.
Another contribution of our work is to augment the training data provided by the existing code by generating negative training examples through simple code transformations.






Name-based bug detection relates to existing learning-based approaches that consider identifiers.
JSNice~\cite{Raychev2015} and JSNaughty~\cite{Vasilescu2017} address the problem of recovering meaningful identifiers from minified code, using conditional random fields and statistical machine translation, respectively.
Another approach summarizes code into descriptive names that can serve, e.g., as method names~\cite{Allamanis2016}.
The Naturalize tool suggests more natural identifiers based on an n-gram model~\cite{Allamanis2014}.
Applying such a tool before running our name-based bug detectors is likely to improve its effectiveness.

A recent graph-based learning approach~\cite{Allamanis2017b} applies gated graph neural networks to an augmented AST representation of code, and shows that a learned model can predict how to name a variable and which variable to use in a given context.
To learn representations of code, their approach relies on static type information, which is not available in our setting.


Word embeddings, e.g., Word2Vec approach~\cite{Mikolov2013a}, are widely used in natural language processing, which has inspired our embeddings of identifiers.
Other recent work has proposed embeddings for source code, e.g., for API methods~\cite{Nguyen2017}, for terms that occur both in programs and natural language documentation~\cite{Ye2016}, as well as a generic, AST-based representation of code~\cite{Alon2018}.
Our evaluation shows that a learned embedding is beneficial for the problem of bug detection.
Incorporating another embedding of identifiers than our Word2Vec-based embedding into our framework is straightforward.


\subsection{Machine Learning and Language Models for Analyzing Bugs}

Even though it has been noted that buggy code stands out compared to 
non-buggy code~\cite{Ray2016}, little work exists on automatically detecting bugs via machine 
learning.
Murali et al.\ train a recurrent neural network that probabilistically models sequences of API calls and then use it for finding incorrect API usages~\cite{Murali2017}.
In contrast to \name{}, their model learns from positive examples only and focuses on bug patterns that can be expressed via probabilistic automata.
Bugram detects bugs based on an n-gram model of code.
Similar to the above, it learns only from positive examples.

Choi et al.~\cite{Choi2017} train a memory neural network~\cite{Weston2014} to classify whether code may produce a buffer overrun.
The VulDeePecker approach~\cite{Li2018} trains a neural network to determine whether a code snippet suffers from specific kinds of vulnerabilities.
Their models learn from positive and negative examples, but the examples are manually created and labeled.
A key insight of our work is that simple code transformations that seed artificial bugs provide many negative examples, and that learning from these examples yields a classifier that is effective for real-world code.

Finally, Wang et al.\ use a deep belief network to find a vector representation of ASTs, which are used for defect prediction~\cite{Wang2016a}.
Their approach marks entire files as likely to (not) contain a bug. However, in contrast to our and other bug finding work, their approach does not pinpoint the buggy location.






\subsection{Bug Finding}

A related line of research is specification mining~\cite{Ammons2002} and the use of 
mined specifications to detect unusual and possibly incorrect code~\cite{Wasylkowski2009,icse2012-statically,Liang2016}.
In contrast to our work, these approaches learn only from correct 
examples~\cite{Hangal2002} and then flag any code that is unusual compared 
to the correct examples as possibly incorrect, or search for inconsistencies 
within a program~\cite{Engler2001}.
Our work replaces the manual effort of creating and tuning such approaches by learning and does so effectively by learning from both correct and buggy code examples.


Our name-based bug detectors are motivated by manually created name-based program analyses~\cite{issta2011,icse2016-names,oopsla2017}.
The ``swapped arguments'' bug detector is inspired by the success of a manually developed and tuned name-based bug detector for this kind of bug~\cite{oopsla2017}.
For the other three bug detectors, we are not aware of any existing approach that addresses these problems based on identifiers.
Our work differs from existing name-based bug detectors (i) by
exploiting semantic similarities that may not be obvious to a lexical comparison of identifiers,
(ii) by replacing manually tuned heuristics to improve precision and recall with automated learning from examples,
and (iii) by applying name-based bug detection to a dynamically typed language.


\subsection{Other Related Work}

Our idea to create artificial, negative examples to train a binary classifier can be seen as a variant of noise-contrastive estimation~\cite{gutmann2010noise}.
The novelty is to apply this approach to programs and to show that it yields effective bug detectors.
The code transformations we use to create negative training examples relate to mutation operators~\cite{jia2011analysis}.
Additional mutation operators for name-related bugs, e.g., mined from version histories~\cite{Brown2017a}, could be an interesting way to extend \name{} with further bug detectors.

To evaluate bug detection tools, automated bug seeding has been proposed.
LAVA~\cite{Dolan-Gavitt2016} and EvilCoder~\cite{Pewny2016} introduce artificial security bugs into code to evaluate vulnerability detection tools.
Their work differs from ours by focusing on vulnerabilities instead of name-related bugs and by seeding bugs for a different purpose.



\section{Conclusions}

This paper addresses the problem of name-based bug detection through a machine learning-based approach.
Our \name{} framework learns bug detectors that distinguish correct from incorrect code by reasoning about the identifier names involved in the code.
The key insights that enable the approach are (i) that reasoning about identifiers based on a learned, semantic representation of names is beneficial and (ii) that artificially seeding bugs through simple code transformations yields accurate bug detectors that are effective also for real-world bugs.
In contrast to previous work on name-based bug detection, we target a dynamically typed language, where names convey information about the intended semantics of code, which is valuable in the absence of static types.
Applying our framework and three bug detectors built on top of it to a large corpus of code shows that the bug detectors have an accuracy between 89\% and 95\%, and that they detect 102 programming mistakes with a true positive rate of 68\%.
In the long term, we envision our work to complement manually designed bug detectors by learning from existing code and by replacing some of the human effort required to create bug detectors with computational effort.

\begin{acks}
This work was supported in part by the German Research Foundation within the Emmy Noether project ConcSys and the Perf4JS project, by the German Federal Ministry of Education and Research and by the Hessian Ministry of Science and the Arts within CRISP, by the Hessian LOEWE initiative within the Software-Factory 4.0 project, by NSF grants CCF-1409872 and CCF-1423645, and by a gift from Fujitsu Laboratories of America, Inc.
\end{acks}

\bibliographystyle{ACM-Reference-Format}
\bibliography{references}
\end{document}